# Topotactic Transition: A Promising Opportunity for Creating New Oxides

*Ziang Meng, Han Yan, Peixin Qin, Xiaorong Zhou, Xiaoning Wang, Hongyu Chen, Li Liu, Zhiqi Liu\**

School of Materials Science and Engineering, Beihang University, Beijing 100191, China
Email: zhiqi@buaa.edu.cn

**Topotactic transition is a structural phase change in a matrix crystal lattice mediated by the ordered loss/gain and rearrangement of atoms, leading to unusual coordination environments and metal atoms with rare valent states. As early as in 1990s, low temperature hydride reduction was utilized to realize the topotactic transition. Since then, topological transformations have been developed via multiple approaches. Especially, the recent discovery of the Ni-based superconductivity in infinite-layer nickelates has greatly boosted the topotactic transition mean to synthesizing new oxides for exploring exotic functional properties. In this review, we have provided a detailed and generalized introduction to oxygen-related topotactic transition. The main body of our review include four parts: the structure-facilitated effects, the mechanism of the topotactic transition, some examples of topotactic transition methods adopted in different metal oxides (V, Mn, Fe, Co, Ni) and the related applications. This work is to provide timely and thorough strategies to successfully realize topotactic transitions for researchers who are eager to create new oxide phases or new oxide materials with desired functions.**

# 1. Introduction

Transition metal oxides involves a pretty wide range of material categories, among which 3$d$ metal oxides have attracted great research attention owing to their abundant physical properties. They exhibit a large variety of unusual characteristics and have been extensively applied in batteries, supercapacitors, oxide electronic devices, chemical catalysis reaction and solar cells.[1]-[14]

On the other hand, topotactic phase transition has been adopted to create new phases or new materials for oxides. Specifically, a topotactic transition refers to a structural phase change accomplished by the ordered loss/gain and rearrangement of atoms at specific positions in a lattice while remaining the same crystallographic frameworks with the parental phase. Compared with the soft-chemical synthesis that is usually considered as a reduction process bounded with binary metal hydrides as reducing agents,[15] the topotactic transition involves a broader scope, which can be achieved through many methods including but not limited to post-annealing, redox reaction, electrochemical process, electron beam irradiation and proton intercalation. Especially, after the recent sensational discovery of superconducting infinite-layer nickelates $Nd_{0.8}Sr_{0.2}NiO_2$ (NSNO) which has been obtained via the $CaH_2$ topotactic reduction from perovskite (PV) $Nd_{0.8}Sr_{0.2}NiO_3$,[16] the topotactic transition approach to engineering new oxides has gained great attention.

In contrast to ordinary thermodynamics-driven synthesis processes, topotactic transitions exhibit a kinetics-driven nature, and could usually occur at relatively low temperatures. For example, the superconducting infinite-layer NSNO thin film was obtained at 260-280 °C, [16] which is much lower than normal solid-state reaction synthesis temperatures that are typically around 1000 °C.[17] More importantly, due to the low synthesis temperature of the topotactic transitions, metastable metal oxides which are inaccessible at high temperatures could be stabilized.

It is kind of unique that topotactic transitions are fundamentally different from usual chemical dopings as they are based on a long-range ordered structure phase change with a topotactic nature. Therefore, it is an effective method to creating new oxide materials or new oxide phases. In this way, one could be able to unveil some completely unprecedented properties via fabricating new materials.

In this review, we mainly focus on oxygen-related topotactic transitions in different metal oxides, which have been studied for decades yet surprising discoveries still emerge in an endless stream. The whole review contains the following parts: crystallographic structures that could facilitate the topotactic transition, kinetics-driven mechanisms for oxygen diffusion, experimental realization of the topotactic transition for five types of common metal oxides, potential material and device applications, and finally a brief outlook.

## 2. Favorable crystallographic structures

The realization of an ordered rearrangement of oxygen vacancies/atoms could be largely distinct for different crystal structures of metal oxides. Therefore, the crystallographic structure is especially important as the structural features of a precursor, intermediate phases and products determine whether the topotactic transition can occur. In this part we will have a brief overview on different structures which may facilitate the topotactic transition.

### 2.1. PV and layered PV structures

The PV and layered PV phases are the most common structures for realizing the topotactic transition. The PV structure shares a simple $ABO_3$ formula, where the B site lies in the center of an octahedron oxygen anions and the A site locates in the center of a unit cell which is built from a corner-sharing $BO_6$ octahedron.[2],[18] The latter layered PV structures are intergrowths of PV and other structures, which are formed by interlacing two-dimensional PV planes with other cationic structures including the Dion-Jacobson phase, the Ruddlesden-Popper (RP) phase and so on.[18] The versatility of these structures is caused by the structural flexibility since the rotation of the corner-shared octahedra may reduce the diffusion energy and accommodate different ionic size.[19] Their layered structures also make them prone to undergo interlayer transformations without breaking covalent bonds within the layers.[20] This also explains why the layered structure of PV or PV-related phases can facilitate the topotactic oxygen migration. More specifically, on the one hand, the anisotropic atomic bonds in these layered structures naturally amplify the diversity of atomic mobility, increasing the migration rates of oxygen. On the other hand, the layered structures can cause the difference of chemical environment at distinct sites, which results in the variation of stability and brings certain selectivity for the transition.[21]

**2.2. Brownmillerite (BM) structures**

Brownmillerites are oxygen deficient layered perovskites of the general formula $A_2B_2O_5$ having alternating layers of oxygen octahedra and tetrahedra. They belong to a classical structure of intermediate phases or products of the topotactic transition. For example, during the transition from $SrFeO_3$ to $SrFeO_2$, the appearance of the BM phase $SrFeO_{2.5}$ is important for revealing the topotactic transition mechanism, which we will elaborate in the next part.[22] The BM phase has a one-dimensional oxygen vacancy channel along the (110) orientation parallel to the tetrahedra chains,[23] as can be seen in **Figure 1**a. These tetrahedra chains can cooperatively twist to result in the left or right hand symmetries. Furthermore, the different arrangements of the twisted chains can generate different distortion patterns, leading to various space groups such as I2bm, Pbcm and Pnma, shown in **Figure 1**b.[24] Not only can the ordered oxygen vacancy channel facilitate the ionic $O^{2-}$ migration in a great extent,[25] but theoretical calculations conducted by Stølen *et al.* also proves that the low-energy oxygen connective schemes within the tetrahedra layer is vital for its high ionic conductivity.[26] Additionally, the multiform collaborative distortions of the octahedra and tetrahedra may further reduce the energy barrier for oxygen ion migration.

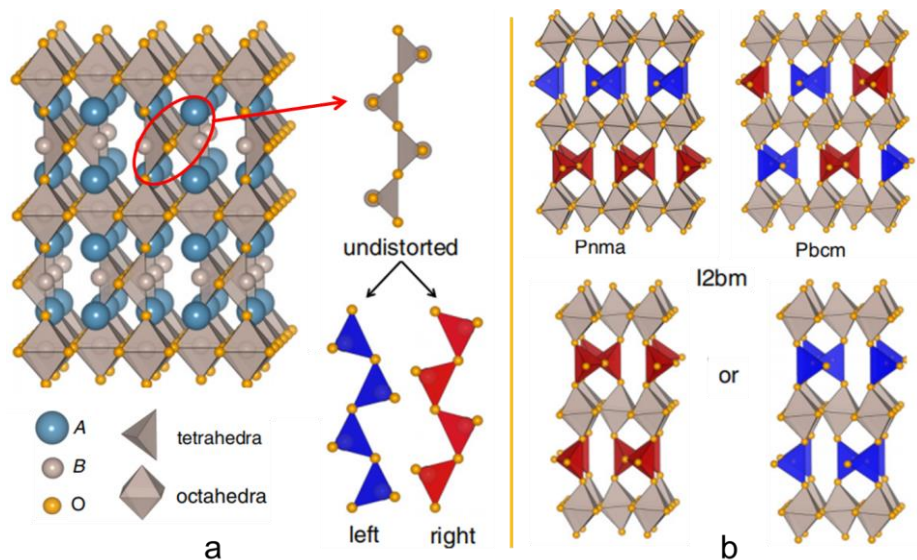

**Figure 1.** The structure of a BM phase with different space groups. a) The crystal structure of the BM phase and the two types of tetrahedra chains marked by red and blue. b) The structure of the BM phase belongs to Pnma, Pbcm and I2bm space groups, respectively. Reproduced with permission.[24] Copyright 2015, American Physical Society.

It is worth emphasizing that that the oxygen diffusion process in a $A_2B_2O_5$ system is rather complicated. That is because this single formula could correspond to several different structures and the oxygen diffusion in different structures could be largely distinct. For example, previous studies suggest that the oxygen diffusion process could be influenced by local crystal symmetries as well as the atoms located at A and B sites.[27],[28]

## 2.3. Infinite-layer structures

In nature, infinite-layer oxides with a chemical formula $ABO_2$ have been rather rare as most of these forms of materials are metastable, probably due to unusual coordination and ultralow chemical valence states of cations. Such a structure is typically obtained by the topotactic transition from the chemical reduction of the BM or PV phases.[29] In turn, it can be topotactically converted to the BM or PV phases by adding oxygen atoms via oxygen post annealing. For instance, it was found that the in the infinite-layer $LaNiO_2$ can be fully changed to the PV $LaNiO_3$ by an annealing under an oxygen flow at 680°C for 12 h.[30]

Compared with the stoichiometric PV phase and the BM phase that contains one-dimensional oxygen vacancy channels, the infinite-layer oxides are full of two-dimensional oxygen vacancy planes, which could serve as active oxygen diffusion platforms so that the reversible topotactic transition could be easily allowed. Therefore, the metastable nature of the infinite-layer structures ensures that they are favorable for the topotactic transition to more stable BM and PV phases.

## 3. Mechanism of oxygen-related topotactic transition

The key mechanism of the topotactic transition is related to the diffusion properties of ions, which involves multiple kinetic processes. The occurrence of the topotactic transition is commonly believed as the differences in the diffusion ability of various ions at relatively low temperatures. Ions with relatively small radius such as $O^{2-}$, $F^-$, $Li^+$ and $Na^+$, have higher mobility, thus making them easier to insert or remove from the crystal lattice framework.

As emphasized in the previous reports on the infinite-layer $SrFeO_2$,[31] the topotactic transition is a kinetically controlled reaction, and the final product is not always the most stable one while it is the one that forms first.[32] Understanding the dynamics of the oxygen ion migration is a prerequisite for unveiling the mystery of the topotactic transition. With the advancement of time-resolve and *in-situ* detection methods such as *in-situ* electron microscopy,[33]-[36] *in-situ* neutron diffraction,[37] and precise thermogravimetric analysis.[38]

### 3.1. Two important points worth emphasizing

The first is a common misconception that may exist for the oxygen-related topotactic transition. That is only the oxygen ions in specific sites are moving while others stay static during a topotactic transition. As pointed out in the previous literature,[32] the topotactic transition could be more like a two-step reaction process, which includes the remove of a certain amount of oxygen and the subsequent rearrangement of the entire remaining oxygen framework.[39] A similar conclusion was drawn by Inoue *et al.*.[40] They achieved the topotactic transition to $CaFeO_2$ from two BM-phase $CaFeO_{2.5}$ thin films with different orientations (010) and (001). As shown in **Figure 2**, both two films can be transferred into the same (001)-orientated $CaFeO_2$ with an infinite-layer structure, which indicates that there are two types of oxygen migration patterns. The energy barrier of these two types of oxygen migration is different, leading to a longer 96 h reaction time for Type 2. The similar result was reported by Chen *et al.*,[22] which further supports the overall oxygen atom rearrangement scenario.

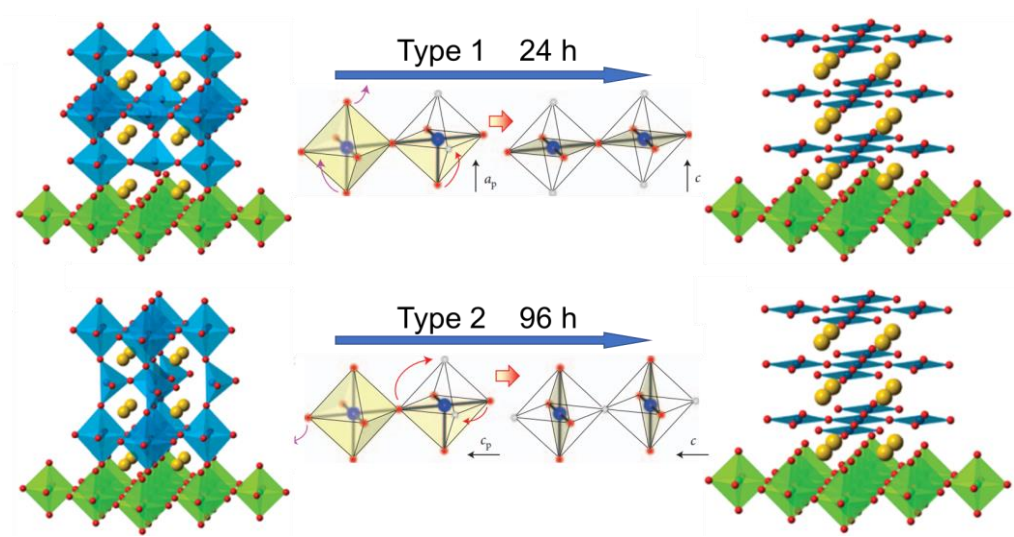

**Figure 2.** The two types of oxygen migration patterns of the topotactic transition from SrFeO$_{2.5}$ to infinite-layer SrFeO$_2$. Reproduced with permission.[39] Copyright 2012, Chemical Society Reviews.

The second is that the ordering of oxygen vacancies during the topotactic transition typically leads to remarkable changes in lattices constants. For example, in 2017, Jang *et al.* reported the *in-situ* observation of the electron-beam-irradiation-induced formation of oxygen vacancies in LaCoO$_3$ using scanning transmission electron microscopy (STEM).[41] The formation of oxygen vacancies can be clearly seen in **Figure 3**a and b. With a longer irradiation time, the two distinct BM regions nucleate from different sides and finally evolve into a single oxygen deficient layer. The whole process is accompanied by an obvious out-of-plane La-La distance variation (**Figure 3**c). The distance between two adjacent La atoms is demonstrated in **Figure 3**d and the structure of topotactic transition via intermediate phase is shown in **Figure 3**e. These results indicate that apart from the ordered arrangement of oxygen anions, the topotactic transition also involves the sizeable adjustment of the cation framework, which may thus result in different ground states and lead to novel physical properties.

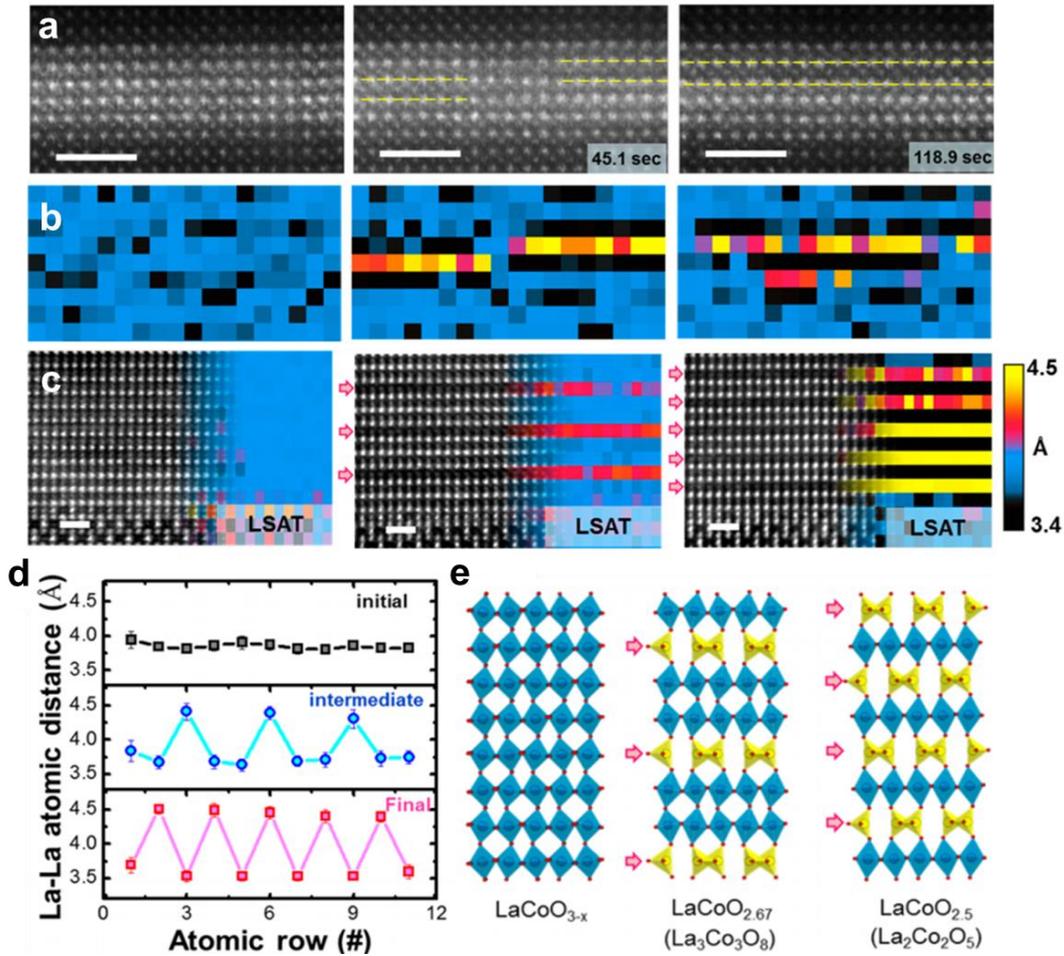

**Figure 3.** Topotactic transition from $LaCoO_{3-x}$ to $LaCoO_{2.5}$ via $LaCoO_{2.67}$. a) ADF images of the forming of electron-beam-induced oxygen vacancy defects and b) corresponding spacing maps of out-of-plane La−La distances. From left to right indicates the longer irradiation time. c) Beam-induced transition in 6 nm $LaCoO_{3-x}$ films on LSAT substrate with corresponding La−La spacing maps. d) Profiles of La−La distance maps in different stages of beam exposure, corresponding the three STEM image in (c). e) Diagrammatic sketch of the topotactic transition in $LaCoO_x$ system. Reproduced with permission.[41] Copyright 2017, American Chemical Society.

**3.2. Oxygen migration mechanism for different methods**

The oxygen-related topotactic transition can be triggered by numerous methods such as an oxygen getter layer, annealing, and a bias voltage. In this part, we will elaborate the oxygen migration mechanism for each approach.

As reported in 2011, an oxygen deficient $SrTiO_3$ (STO) layer on top of a precursor film can be regarded as an oxygen getter, which leads to the topotactic transition of $La_{0.7}Sr_{0.3}MnO_3$ towards $La_{0.7}Sr_{0.3}MnO_{2.5}$.[42] It was elucidated that the phase transition is ensured by the oxygen diffusion from the buried film into the getter layer and the driven force of the transition is the difference in oxygen affinities of the two oxide layers. A similar method was reported earlier in 2008,[43] where a PV-LaAlO$_3$ overlayer was grown to achieve carrier doping in the underlying TiO$_2$ thin film.

The work carried by Hu *et al.* used the scanning transmission electron microscopy with *in-situ* heating capability to observe the structural evolution of the LaSrMnO$_x$ (LSMO) thin film during annealing in high vacuum (10$^{-9}$ Torr) at 500 °C under different strain states.[44] A fully transition from PV-LSMO to BM-LSMO has been detected and interestingly, the whole process is proved to be reversible. However, the oxygen diffusion process and the related topotactic transition is strongly strain dependent. (**Figure 4**). Similar strain-influenced topotactic transition from LaCoO$_{2.67}$ to LaCoO$_{2.5}$ was reported by Zhang *et al.* in 2021.[45]

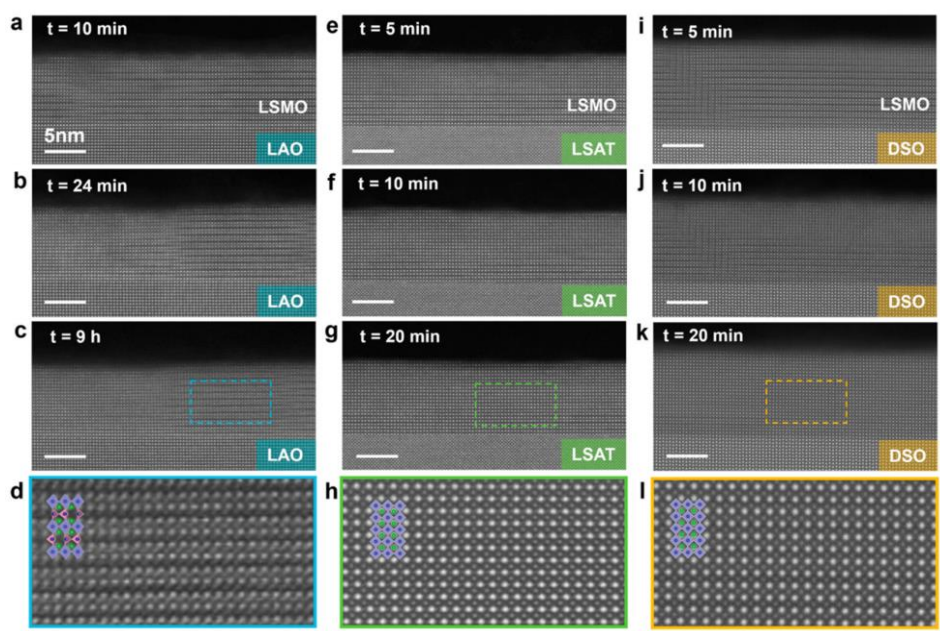

**Figure 4.** Strain dependent reverse transition from BM-LSMO to PV-LSMO under ultra-high vacuum with different substrate after the temperature increasing back to room temperature. a)-d) STEM image shows the slow oxidation of BM-LSMO on large compressive strain LAO substrate for 10 min, 24 min and 9 h respectively. d) The enlarged image of the remaining BM phase marked by blue area in (c). e)-h) and i)-l) STEM image shows the rapid oxidation of BM-LSMO on strain moderate LSAT and tensile DSO substrate for 5 min, 10 min and 20 min respectively. h) and l) The enlarged oxidized PV phase marked by green and yellow area in (g) and (k). Reproduced with permission.[44] Copyright 2022, Elsevier.

Reversible topotactic phase transition triggered by electric voltage in resistive switching devices is worth studying as well. In 2021, Mou *et al.* developed this kind of devices based on the SrCoO$_x$ (SCO) system with different orientate. The voltage induced topotactic transition from insulating BM-SCO to conductive PV-SCO was achieved.[46] As shown in **Figure 5**, the oxygen ions at the bottom SrRuO$_3$ (SRO) electrode is driven by a negative bias towards the top Au electrode, thus forming a conducting filament. During this process, at the filament region, the insulating BM-SCO is topotactically transited to conductive PV-SCO, leading to a low resistance state. On the contrary, when applying a reverse positive bias to the bottom SRO layer, oxygen ions are attracted back, and the conductive PV-SCO changes to insulating BM-SCO, resulting in a high resistance state. Additionally, the energy of two types of oxygen migration passes within the oxygen tetrahedral layers and across the oxygen tetrahedral layers has been calculated. The migration routine alone the oxygen vacancy channel (vertical red arrow in **Figure 5**e) has the lowest energy barrier.

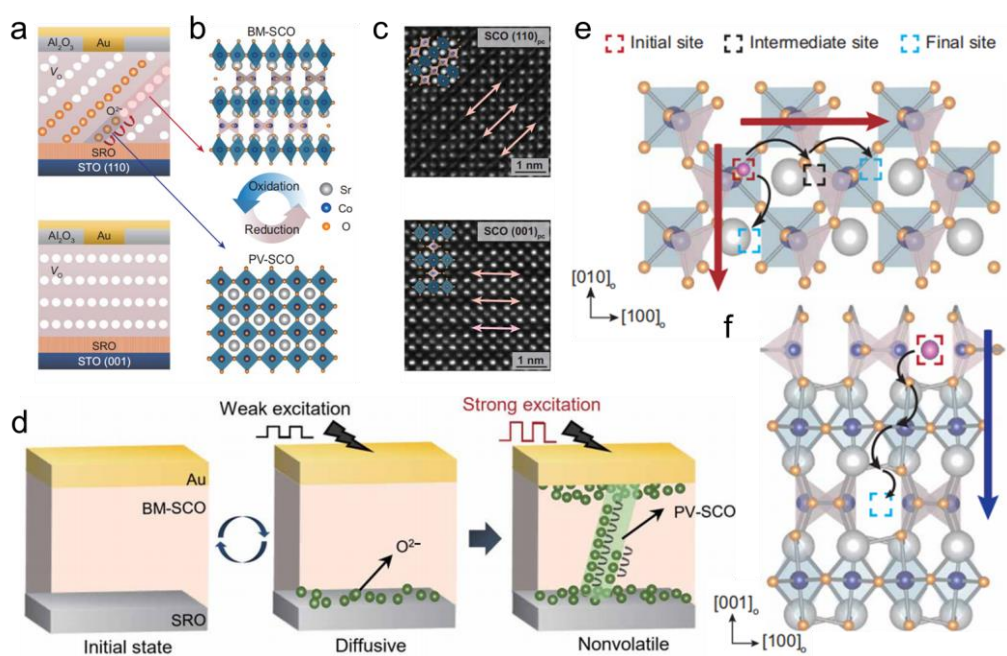

**Figure 5.** Structural design and switching mechanism of a resistive switching device. a) Schematics of RAM based on (110) and (001) orientations of SCO. b) Reversible voltage-induced topotactic transition between BM-SCO and PV-SCO. c) STEM image of (110) and (001) oriented BM-SCO. The pink arrow indicates the oxygen tetrahedral layer. The index shows the specific structure models. d) Schematic of the memristor switching mechanism. The oxygen migration pathways e) within the oxygen tetrahedral layers and f) across the oxygen tetrahedral layers. Reproduced with permission.[46] Copyright 2021, American Association for the Advancement of Science.

### 3.3. Complexity of the topotactic transition kinetics

Although many studies have been carried to identify the mechanism of topotactic transition with different reacting conditions, it is still far from enough. In 2006, Toquin *et al.* conducted the time-resolved *in-situ* studies of oxygen intercalation through neutron diffraction and X-ray absorption spectroscopy.[37] They characterized the two intermediate phases, $SrCoO_{2.75}$ and $SrCoO_{2.82\pm0.07}$, during the transition from BM-SCO to PV-SCO. When BM-SCO is oxidized, the native one-dimensional oxygen vacancy channels in the $CoO_4$ tetrahedral layers are filled up gradually, resulting in ordered intermediate phases. This involves a transition from four-fold tetrahedral coordination to five-fold square pyramidal coordination, and finally to six-fold octahedral coordination. They firstly reported the formation of $O^-$ accompanying by the valence state fluctuation of Co during the phase transition. They believed that the small radius of $O^-$ promotes the topotactic rearrangement process by reducing the potential energy. It was pointed out that the conventional assumption that good ionic conductors favor structural disorders is dubious when dealing with the topotactic transition with great complexity.

With the help of more advanced *in-situ* observation techniques, more studies on the mechanism of the topotactic transition are needed, as most previous studies have only observed the formation of ordered oxygen vacancy structures, but cannot identify the specific movement paths of individual oxygen atoms during the formation of these vacancies. The method of isotope labeling may be able to explore the dynamic movement process of individual oxygen atoms, thus providing a straightforward study on the kinetics of the topotactic transition.

**4. Examples of topotactic transition of metal oxides**

**4.1. V-Based Oxides**

As a result of the quasi-room-temperature metal-insulator transition, vanadium dioxide ($VO_2$) is a promising material for next-generation optical and electronic devices as it exhibits giant changes in resistivity and infrared transmittance across the transition.[47]-[51]. An interesting alternative approach to obtaining high-quality $VO_2$ thin films is to utilize the topotactic oxidation of a $V_2O_3$ precursor. For example, Yamaguchi *et al.* reported the synthesis of $VO_2$ thin film via a post-epitaxial topotaxy method.[52] They managed to topotactically oxidize (0001)-oriented $V_2O_3$ films into (010)-oriented $VO_2$ thin films on $Al_2O_3$ substrates. Similarly, Okimura and Suzuki[53] realized the successful topotactic transition of $V_2O_3$ films into homogeneous crystalline $VO_2$ films via oxygen post annealing. The optimal annealing temperature for yielding the pure $VO_2$ phase was found to be ~450 °C at $pO_2$ = 500 Pa. Higher annealing temperatures will cause the over oxidation towards the formation of $V_3O_7$ and $V_2O_5$ phases. Similar topotactic transitions in $V_2O_3$ were also achieved by Nomoto *et al.* with Al-doped ZnO (AZO) seed layer[54] and Matamura *et al.* by annealing in $O_2$ atmosphere.[55] The atom migration process from $V_2O_3$ to $VO_2$ is detailed depicted in **Figure 6**a and the x-ray diffraction (XRD) associated with reciprocal space maps is shown in **Figure 6**b-e. A common point of these studies is that the crystal structures of both $V_2O_3$ and $VO_2$ appeared to share a nearly identical NiAs-type structural framework.

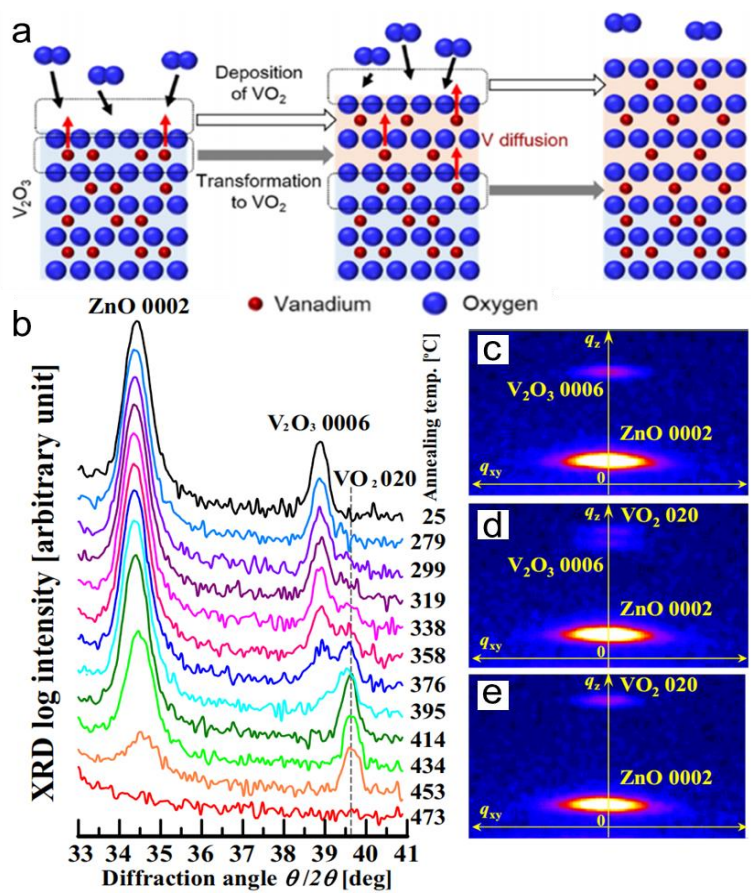

**Figure 6.** Phase transition from $V_2O_3$ to $VO_2$. a) The theoretical transformation process by $O_2$ oxidation. The pink regions in the second and last schematics represent $VO_2$. Reproduced with permission.[55] Copyright 2022, America Chemical Society. b) A serious of XRD patterns of $V_2O_3$ films on the AZO seed layer with different annealing temperature. Reciprocal space maps of $V_2O_3$ films at Temperature of c) 25, d) 376, and e) 414 °C. Reproduced with permission.[54] Copyright 2020, Elsevier.

Besides $VO_2$, in 2000, Gopalakrishnan's group reported the transition of the Ruddlesden-Popper (RP) phase oxide $K_2La_2Ti_3O_{10}$, to $(Bi_2O_2)La_2Ti_3O_{10}$, $MLa_2Ti_3O_{10}$ (M = Pb, Ba, Sr) and $(VO)La_2Ti_3O_{10}$, by a novel metathesis reaction with BiOCl, $MCl_2$ and $VOSO_4·3H_2O$, respectively.[56] The reaction towards $(Bi_2O_2)La_2Ti_3O_{10}$ and $MLa_2Ti_3O_{10}$ required high temperatures above 600 °C. However, it was found that the reaction towards $(VO)La_2Ti_3O_{10}$ occurred at only around 100 °C via a topotactic route.

In 2016, Hyojin Yoon's group focused on the reversible phase transition among insulating $VO_2$, metallic $H_xVO_2$, and insulating $HVO_2$.[57] $VO_2$ could be an ideal model system for hydrogen incorporation as presented in **Figure 7**a. Upon the topotactic introduction of H, the electrical properties of $VO_2$ can be greatly affected.[58],[59] In order to reduce the hydrogen dissociation energy, the Pt-catalyst-assisted low-temperature hydrogenation process was adopted (**Figure 7**b). High-resolution transmission electron microscopy images before and after the hydrogenation of $VO_2$ is shown in **Figure 7**c,d. Without disrupting the rutile crystal structure, the hydrogen incorporation is successfully achieved.

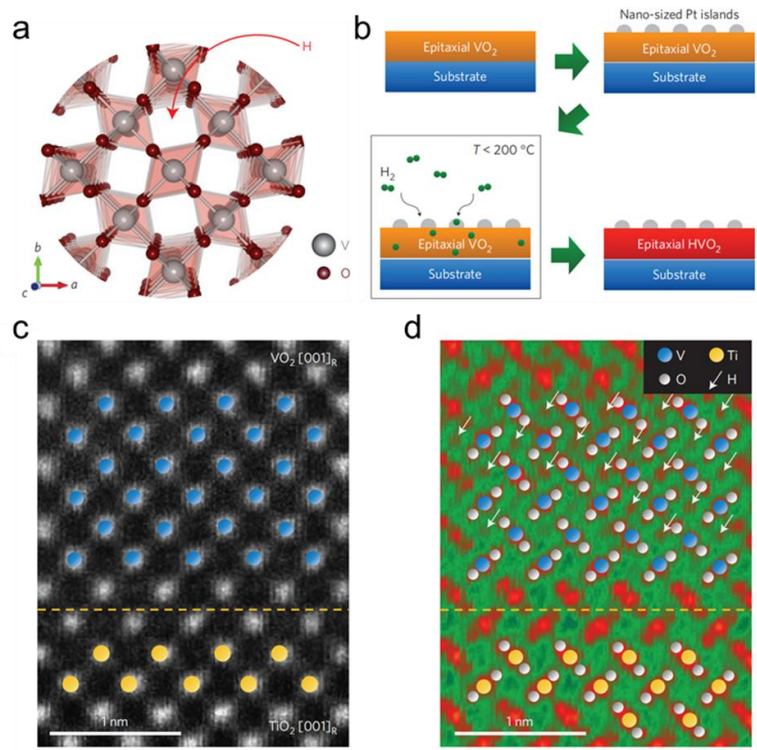

**Figure 7.** Low-temperature hydrogenation of VO$_2$. a) c-axis empty H channel in the rutile VO$_2$ lattice. b) Diagram of whole hydrogenation process of VO$_2$ thin films with hydrogen spillover method. c) High-angle annular dark field scanning transmission electron microscopy (HAADF-STEM) and d) Annular bright field (ABF-STEM) image of the fully hydrogenated HVO$_2$ epitaxial film grown on TiO$_2$ substrates. Yellow dashed line indicated the interface between thin film and substrate. Reproduced with permission.[57] Copyright 2016, Springer Nature.

### 4.2. Mn-Based Oxides

Manganates are famous for the colossal magnetoresistance effect.[60] Although Park's group at Yonsei University failed to realize a uniform topotactic transition for LSMO in 2016,[61] the substantial progress for the topotactic reduction of LSMO was later achieved by Cao *et al*. in 2019.[62] Specifically, they accomplished the oxygen-related reversible topotactic interconversions between the PV-LSMO and the BM-LSMO by either vacuum or oxygen post annealing. The ordered loss of oxygen atoms triggered by vacuum annealing is evidenced by *in-situ* XRD spectra shown in **Figure 8**, demonstrating a topotactic structural transition.

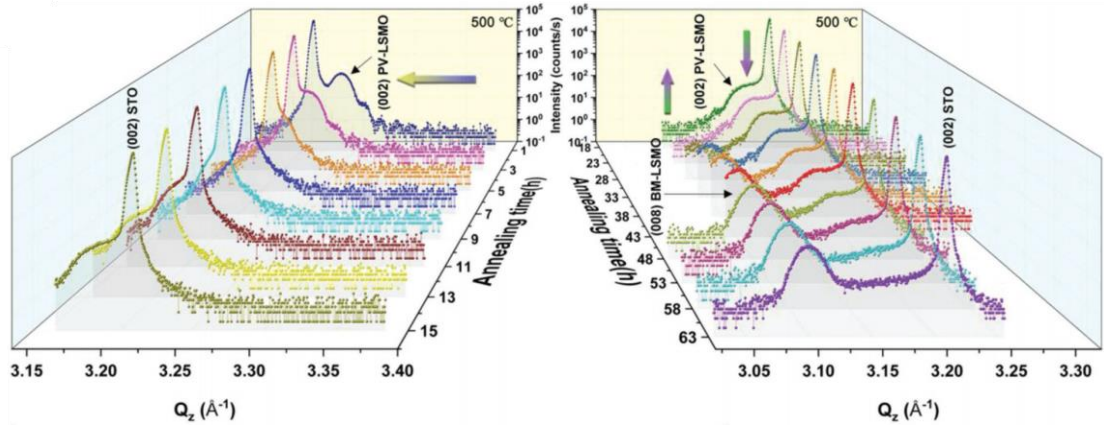

**Figure 8.** *In-situ* XRD scans of the 45-nm-thick $La_{0.7}Sr_{0.3}MnO_{3-\delta}$ film on a STO substrate in 500 ℃ with increasing annealing time. Reproduced with permission.[62] Copyright 2019, Wiley-VCH.

More interestingly, Liu *et al*. have realized the topotactic transition for LSMO films[63] via three different methods including $CaH_2$ reduction and *in-situ* oxygen vacancy diffusion via an oxygen getter layer. For the first approach, they managed to obtain the BM phase by burring samples into $CaH_2$ powders, and annealing them at 350-450 °C for 12 h. It is worth emphasizing that a capping layer seems rather important for stabilizing the BM phase in air. Otherwise, the topotactically reduced BM phase disappears upon three days in air. Regarding their second topotactic transition method, it was found a 20-nm-thick PV-LSMO thin film can be completely transformed into the BM phase with a 3-nm-thick oxygen deficient STO capping layer. In this work, they also tried the synthesis of BM-LSMO thin films directly at low oxygen pressures. Nevertheless, the BM phase was not obtained.

In addition to the injection/depletion of oxygen atoms in LSMO films, the effective adding of hydrogen atoms could also lead to the successful realization of topotactic transition. For example, in 2022, Mazza *et al*.[64] introduced hydrogen atoms into LSMO films by the Pt catalyst and finally achieved the BM phase via the topotactic transition. Intriguingly, an infinite-layer-like phase was also observed, but unfortunately could not be separated for more detailed characterizations.

Apart from the above topotactic reduction methods, other novel paths to inducing the topotactic transition for LSMO have been reporeted as well. Specifically, Yao *et al.*[65] achieved the PV-BM-PV topotactic phase transitions via the electron beam irradation. As shown in **Figure 9**a,d, oxygen atoms are visible in the original PV phase and are orderly arranged in both the marked A and B layers. After 20 min of electron beam irradiation, oxygen atoms dsappeared in the B layers, resulting in the tetrahedral coordination of Mn (**Figure 9**b,e) and the appearance of BM phase. While the electron beam irradiation was kept for longer than 50min (**Figure 9**c,f), an oxygen deficient PV-like phase was formed. This is highly interesting as the oxygen manipualtion capability of the electron beam could be important for other applications as well.

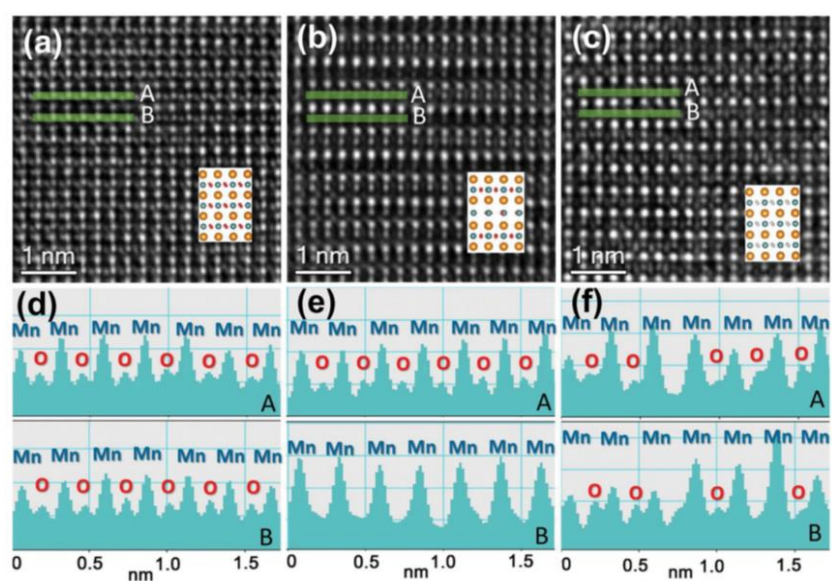

**Figure 9.** Structural characterization of different LSMO phases. HRTEM image of a) PV-LSMO b) BM-LSMO and c) oxygen deficient PV-like LSMO structure phases. The insets show corresponding structural models for each image (La/Sr: yellow, Mn: green, O: red in a) and b), grey in c)). d)-f) Line profiles of two neighboring Mn-O layers indicated by green AB line in the HRTEM images. Reproduced with permission.[65] Copyright 2014, Wiley-VCH.

When electrically injecting hydrogen ions (or protons) into a LSMO thin film, these protons are not able to function as reductants to result in the variation of the oxygen concentration. In turn, they could stay within the oxide lattice via chemical bonding and resultingly form a long range ordered LSMO-H phase, which could exhibit completely distinct physical properties compared with the original phase-separated PV-LSMO. Such a scenario has indeed been experimentally demonstrated by Chen *et al.*[66] as schematized in **Figure 10**. This approach could occur at relatively low temperatures as the main driving force is the electric field.

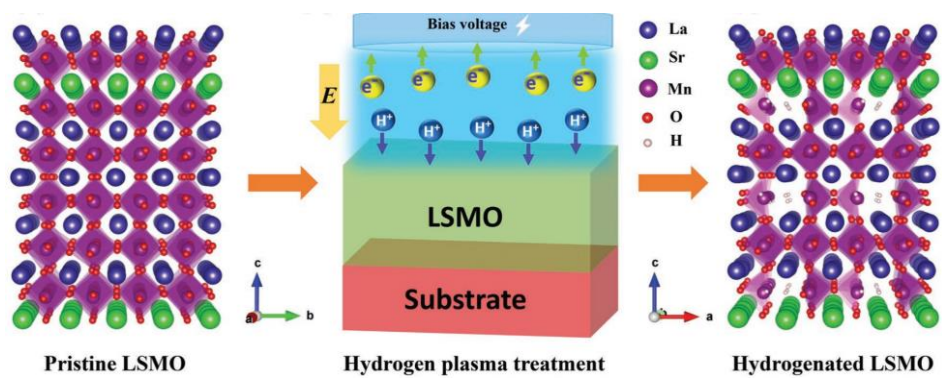

**Figure 10.** Schematics of pristine PV-LSMO films experiencing electrical injection of protons to transform into BM-LSMO. Reproduced with permission.[66] Copyright 2019, Wiley-VCH.

### 4.3. Fe-Based Oxides

Topotactic structural transitions can also be achieved in Fe oxides. In 2007, Tsujimoto *et al.* discovered that the PV-SrFeO$_3$ can be topotactically reduced into the exotic infinite-layer antiferromagnetic SrFeO$_2$ via the reducing agent CaH$_2$.[31] Technically, they sealed the SrFeO$_3$ powder sample with two-molar excess of CaH$_2$ in an evacuated Pyrex tube and reacted at 553 K for two days, resulting in the infinite layer SrFeO$_2$ phase consisting of sheets of apex-linked Fe$^{2+}$O$_4$ square-planes stacked with Sr$^{2+}$ cations as seen in **Figure 11**a,b. The magnetism of SrFeO$_2$ has also been studied which indicated a antiferromagetism order shown in **Figure 11**c,d Similar topotactic transitions were reported by other groups as well.[40],[67],[68] In addition, the SrFeO$_{2.5}$ with a BM phase can also be topotactically reduced into the infinite-layer SrFeO$_2$ phase by CaH$_2$.[69]

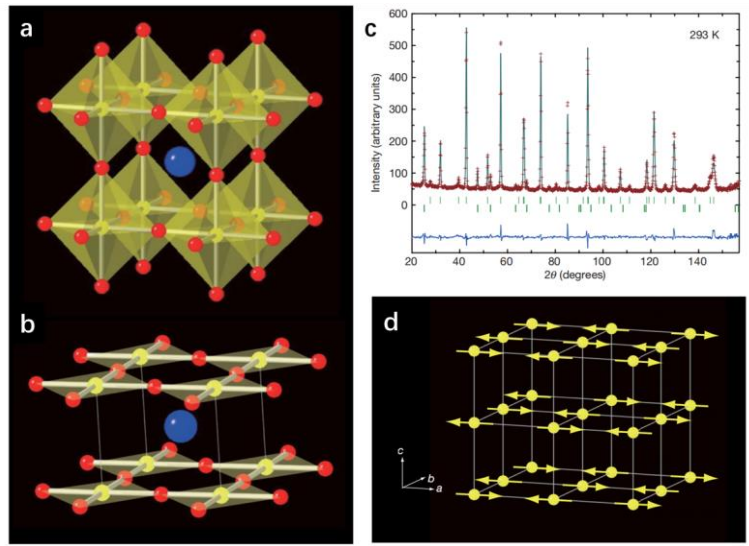

**Figure 11.** The physical characteristics of SrFeO$_3$ and SrFeO$_2$. a) cubic PV-SFO and b) infinite-layer compound SrFeO$_2$. c) The neutron powder diffraction pattern of SrFeO$_2$. The calculated and observed intensities are shown as the solid lines and overlying crosses. The positions of the predicted nuclear (top) and magnetic (bottom) Bragg reflections are indicated by the green lines. The difference between the calculated and observed profiles is plotted in blue at the bottom. d) the magnetic structure of SrFeO$_2$ unit cell, arrows denote the direction of the magnetic moment. Reproduced with permission.[31] Copyright 2007, Springer Nature.

Although the low-oxygen-pressure growth of manganite thin films did not yield the topotactic transition to for the BM phase[63], such an approach has been effective for PV phase Fe oxides. Both Khare *et al.*[70] and Zhao *et al.*[71] have demonstrated the pure BM phase directly fabricated at low oxygen pressure (**Figure 12**).

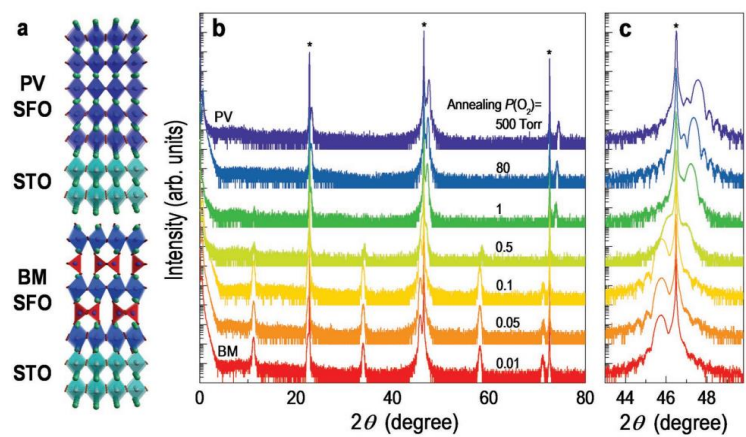

**Figure 12.** Crystal structure of SrFeO$_x$ (SFO) epitaxial thin films. a) Crystal structure of PV-SFO and BM-SrFeO$_{2.5}$ thin films on STO substrate. b) and c) XRD patterns of SFO thin films with different annealing oxygen pressure. Reproduced with permission.[70] Copyright 2017, Wiley-VCH.

Instead of employing chemical reductants or low oxygen pressures, one can also observe the nanoscale local topotactic transition for SFO in well-established resistive switching devices, which consist of a typical metal-oxide-metal capacitor structure. When a SFO thin film is sandwiched between two metal electrodes, the migration of oxygen vacancies upon the external electric field excitations could lead to the reversible PV-BM topotactic phase transitions, accompanied by the giant perpendicular resistance changes as shown in **Figure 13**.[73] Many resistive switching devices have been reported to be based on such a mechanism.[72]-[77]

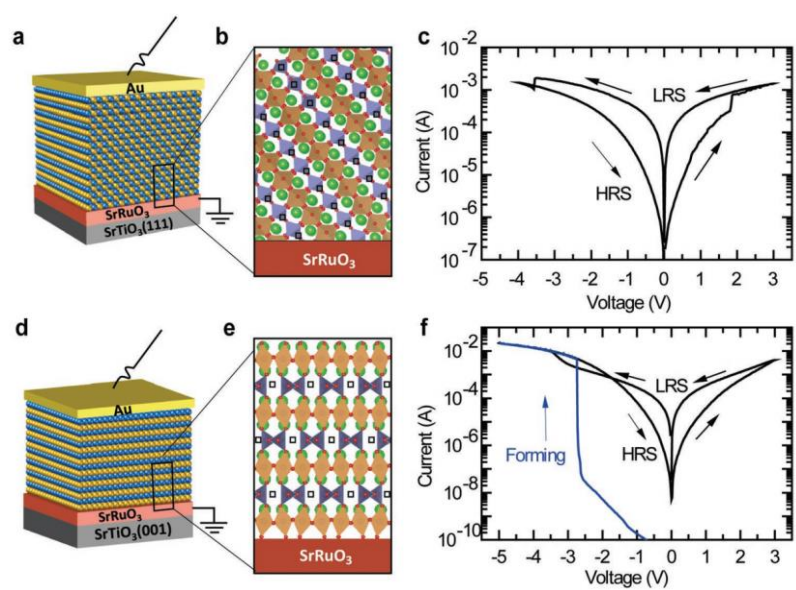

**Figure 13.** RS memory realized in SFO-based layered structure. a) Schematic drawing of Au/SrFeO$_{2.5}$/SRO/STO (111) device with b) out-of-plane ordered oxygen vacancy channels. c) I–V curves of the (111) device. d) Schematic of Au/SrFeO$_{2.5}$/SRO/STO (001) device with e) in-plane ordered oxygen vacancy channels along the tetrahedral layers (blue polyhedral) of the SrFeO$_{2.5}$ layer. f) Forming step (blue) and I–V curve (black) of the (001) device. Purple and orange geometries represent oxygen tetrahedron and octahedron respectively. Reproduced with permission.[73] Copyright 2019, Wiley-VCH.

Similar to the CaH$_2$ reduction, topotactic transitions could also occur at reducing N$_2$ atmosphere for Fe oxides. In 2018, Lee *et al*. reported that a redox-driven reversible topotactic transition in epitaxial SrFe$_{0.8}$Co$_{0.2}$O$_{3-x}$ (SFCO) thin films at low temperatures and atmospheric pressure.[78] It was discovered that complete phase transitions between the PV and BM phases are achieved after heating up to 375 °C in reducing N$_2$ atmosphere. The phase transition is reversible via annealing in N$_2$ or air (**Figure 14**). Besides, a wet-chemical oxidation method is also possible.

For example, Nguyen *et al.* realized a BM to PV topotactic transition for SFCO by NaClO etching.[79]

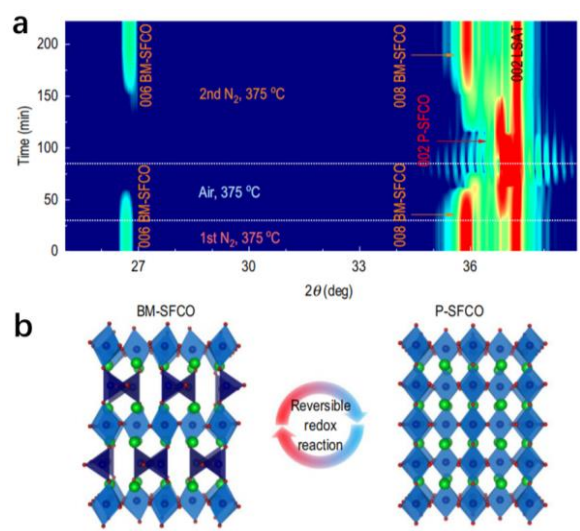

**Figure 14.** Topotactic transition in SFCO system. a) Gas-type-dependent real-time XRD θ-2θ measurement at 375 °C. b) Schematic diagram of reversible redox reaction in SFCO. Reproduced with permission.[78] Copyright 2018, American Physical Society.

### 4.4. Co-Based Oxides

The topotactic phase transitions of Co-based oxides could be relatively easy to achieve via the approaches we mentioned in the above text, for example, $CaH_2$/NaH reduction, vacuum post annealing, low oxygen pressure growth, catalyst-mediated $H_2$ introduction, oxygen getter layer capping and NaClO wet-chemical etching. To be concise, various successful examples of the topotactic transitions in Co-based oxide are summarized in the following table.

| Precursor | Agent | Temp. | Time | Method | Product | Reference |
|---|---|---|---|---|---|---|
| LaSrCoO$_4$ | 8% H$_2$ | 550 °C | 4 h | Soft-chemical | LaSrCoO$_{3.5-x}$ | [80] |
| | NaH | 200 °C | 4 d | | | |
| La$_{0.7}$Sr$_{0.3}$CoO$_3$ | 10$^{-6}$ Pa | 300 °C | 15 min | Anneal | La$_{0.7}$Sr$_{0.3}$CoO$_{2.5}$ | [81] |
| | | 450 °C | | | La$_{1.4}$Sr$_{0.6}$CoO$_4$ | |
| La$_{0.7}$Sr$_{0.3}$CoO$_3$ | 10$^{-24}$ atm | 400 °C | 1 h | Anneal | La$_{0.7}$Sr$_{0.3}$CoO$_{2.5}$ | [82] |
| | 10$^{-12}$ atm | 900 °C | | | La$_{1.4}$Sr$_{0.6}$CoO$_4$ | |
| SrCoO$_{2.5}$ | KOH 500 mV | RT | 180 h | Electrochemical | SrCoO$_3$ | [86] |
| SrCoO$_{2.5}$ | KOH | RT | - | Electrochemical | SrCoO$_3$ | [87] [37] |
| SrCoO$_{2.5}$ | Air 30 mV | 500 °C | - | Electrochemical | SrCoO$_3$ | [88] |
| SrCoO$_{2.5}$ | 4% NACIO | RT | 15 min | Wet chemical etching | SrCoO$_3$ | [89] [92] |
| SrCoO$_{2.5}$ | 600 Torr O$_2$ | 600 °C | 5 min | Anneal | SrCoO$_3$ | [90] |
| SrCoO$_{2.5}$ | 500 Torr O$_2$ | 300 °C | - | Anneal | SrCoO$_3$ | [83] |
| SrCoO$_{2.5}$ | 500 Torr O$_2$ | 200 °C | 5 min | Anneal | SrCoO$_3$ | [84] |
| SrCoO$_{2.5}$ | 5 bar O$_2$ | 350 °C | 10 min | Anneal | SrCoO$_3$ | [85] |
| SrCoO$_3$ | Vacuum | 300 °C | 3 min | Anneal | SrCoO$_{2.5}$ | [91] |
| SrCoO$_3$ | Pt/Ag | RT | 5 h | Noble-Metal-Assisted | SrCoO$_{2.5}$ | [93] [94] |
| | Au | | 8 d | | | |

In this part, we would like to emphasize a recent new approach which has shown positive results for triggering topotactic transitions in La$_{0.5}$Sr$_{0.5}$CoO$_x$ (LSCO). In 2022, Yin *et al.*[95] developed an emergent ionic gel (IG) with a mixture of porous polymer, ionic liquid and acetone solvent. Unlike the usual ionic liquid that needs to be operated at low temperatures, the usage of such an ionic gel guaranteed the all-solid-state operability at room temperature (**Figure 15**a,b). More importantly, it was found that a +2 V gate voltage can result in the topotactic transition of PV-LSCO into BM-LSCO via the migration of oxygen vacancies. Subsequently, the topotactic transition was proved to be reversable while imposing a -3 V gate voltage (**Figure 15**c). Through theoretical calculations, they drew a conclusion that the energy difference between these two phases is quite small (**Figure 15**d), suggesting that the transition can be easily triggered. Additionally, it was found that a compressive strain is favorable for the topotactic transition while a tensile strain is harmful for the structural interconversion with the migration of oxygen vacancies. It is plausible that the strain state of a thin film could affect the structural phase transition kinetics. Yang *et al.*[96] and V. Chaturvedi *et al.*[97] were also achieve the topotactic transition via the adoption of IG.

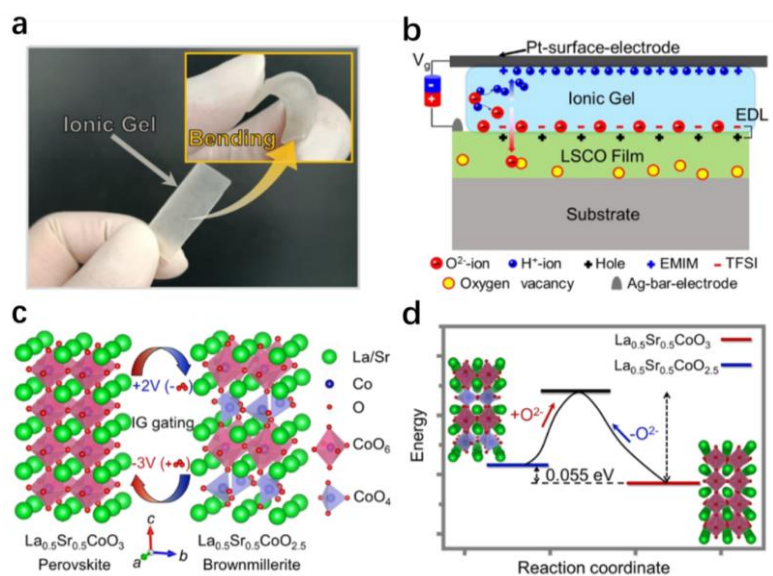

**Figure 15.** The description of IG gating-modulation process. a) Photos of the bending of IG. b) Schematic diagram of all-solid-state IG gating-modulation process. c) Reversible topotactic transition between PV-LSCO and BM-LSCO. d) Energy gap between PV-LSCO and BM-LSCO calculated by theoretical calculations. Reproduced with permission.[95] Copyright 2022, American Physical Society.

Most recently, Woo Jin Kim *et al.* reported an infinite-layer Co-based thin film for the first time.[98] Same as before, they chose $CaH_2$ as the topotactic reduction agent and managed to synthesis single crystal infinite-layer $CaCoO_2$ thin film via 3 hours annealing at 250 °C from BM $CaCoO_{2.5}$ precursor, as shown in **Figure 16**. Unfortunately, the characteristics of superconductivity have not been shown in the infinite-layer cobalt, which is different from infinite-layer nickelate, and means further study is needed.

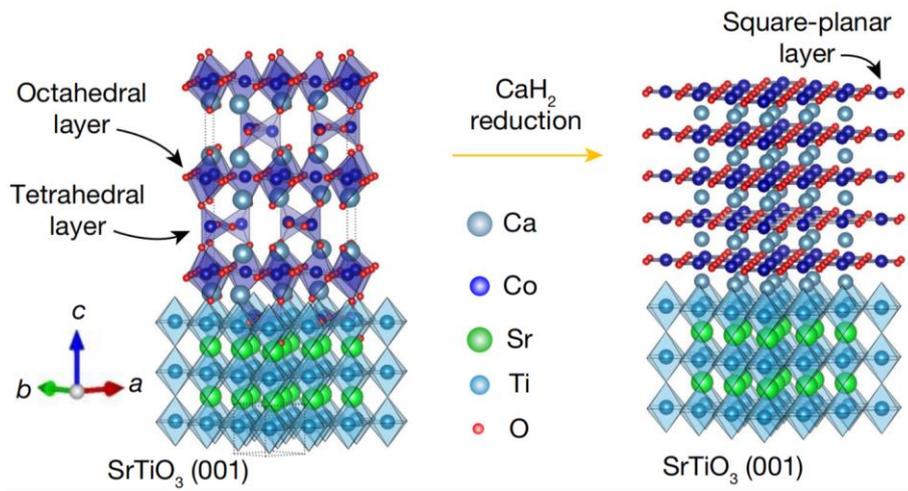

**Figure 16.** Crystal structure of precursor BM CaCoO$_{2.5}$ (left) and product CaCoO$_2$ (right) on (001) oriented STO substrate by the topotactic CaH$_2$ reduction. Reproduced with permission.[98] Copyright 2023, Springer Nature.

### 4.5. Ni-Based Oxides

The most common method for achieving topotactic transitions in Ni-based oxides such as LaNiO$_3$ is chemical reduction. As early as in 1983, Crespin *et al*. reported the successful fabrication of LaNiO$_2$ powders from the LaNiO$_3$ powders via a topotactic reduction in the hydrogen atmosphere[99],[100]. For example, previous studies on the chemical reduction of LaNiO$_3$ are summarized in the follow table.

| Precursor | Agent | Temp. | Time | Environment | Product | Reference |
|---|---|---|---|---|---|---|
| LaNiO$_3$ Powder | Zr metal | 400 °C | 48 h | Evacuated ampoules | LaNiO$_{2.5}$ | [101] [102] |
| LaNiO$_3$ powder | NaH | 190 °C | 3×3 d | Evacuated tube | LaNiO$_2$ | [103] |
| LaNiO$_3$ powder | Excess CaH$_2$ | 300 °C | 24 h | Evacuated Pyrex tube | LaNiO$_2$ | [104] [105] |
| Nd$_{1-x}$Sr$_x$NiO$_3$ powder | 0.25 g CaH$_2$ | 285 °C | 48 h | Evacuated quartz tube | Nd$_{1-x}$Sr$_x$NiO$_2$ | [106] |
| Nd$_{1-x}$Sr$_x$NiO$_3$ powder | CaH$_2$ | 280 °C | 20 h | quartz tube | Nd$_{1-x}$Sr$_x$NiO$_2$ | [107] |
| LaNiO$_3$ Superlattice | 0.5 g CaH$_2$ | 280 °C | 72 h | quartz tubes | LaNiO$_2$ | [108] |
| LaNiO$_3$ film | H$_2$ 1atm | 350 °C | 2-7 h | Flowing H$_2$ | LaNiO$_2$ | [109] |
| LaNiO$_3$ film | 0.25 g CaH$_2$ | 280 °C | 2 h | Pyrex tube | LaNiO$_2$ | [30] |
| LaNiO$_3$ film | CaH$_2$ | 280 °C | 2 h | Glass tube | LaNiO$_2$ | [110] |
| LaNiO$_3$ film | H$_2$ 1atm | 400 °C | 10 min | Flowing H$_2$ | LaNiO$_2$ | [111] |
| LaNiO$_3$ film | 0.5 g CaH$_2$ | 300 °C | 2 h | Quartz tube | LaNiO$_2$ | [112] |

For the past four decades, one of the most important progresses related to the topotactic transitions of nickel-based oxide is the experimental discovery of the Ni-based superconductivity in the infinite-layer nickelates in 2019.[13] Technically, they wrapped each piece of Nd$_{0.8}$Sr$_{0.2}$NiO$_3$/STO film with aluminum foil and sealed such an Al-wrapped package together with 0.1 g of CaH$_2$ powder in a Pyrex glass tube. The tube was heated to 260-280 °C and kept at this temperature for 4-6 h to achieve the infinite-layer NSNO film. The topotactic reaction and the crystal structure are depicted in **Figure 17**.

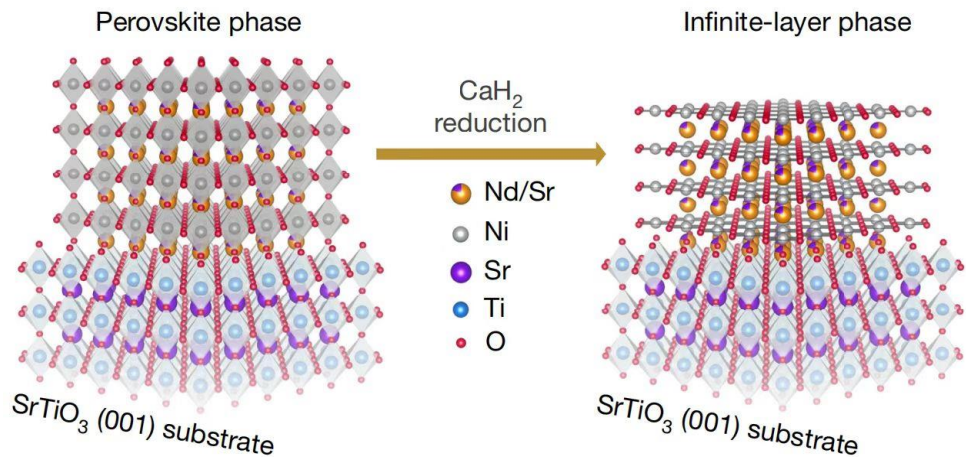

**Figure 17.** Crystal structure of precursor $Nd_{0.8}Sr_{0.2}NiO_3$ (left) and product infinite-layer NSNO (right) on (001) oriented STO substrate by the topotactic $CaH_2$ reduction. Reproduced with permission.[16] Copyright 2019, Springer Nature.

Following this pioneering work, many research groups have managed to synthesize NSNO thin films with $CaH_2$ as a chemical reducing agent. Various reduction conditions that have been adopted by different groups are listed in the following table.

| Precursor | Temp. | Time | Location | Capping layer | Product |
|---|---|---|---|---|---|
| $Nd_{0.8}Sr_{0.2}NiO_3$ | 240°C | 2 h | Pyrex tube | no | $Nd_{0.8}Sr_{0.2}NiO_2$ [113][114] |
|  | 260°C | 6 h |  | 25 nm STO |  |
| $Nd_{0.8}Sr_{0.2}NiO_3$ | 340°C | 100 min | Quartz tube | no | $Nd_{0.8}Sr_{0.2}NiO_2$ [115] |
| $Pr_{0.8}Sr_{0.2}NiO_3$ | 210-240°C | 45-60 min | Pyrex tube | no | $Pr_{0.8}Sr_{0.2}NiO_2$ [116] [120] |
| $Nd_{0.8}Sr_{0.2}NiO_3$ | 340°C | 100 min | Quartz tube | no | $Nd_{0.8}Sr_{0.2}NiO_2$ [117] |
| $Nd_{0.8}Sr_{0.2}NiO_3$ | 340-360°C | 80-120 min | PLD chamber | no | $Nd_{0.8}Sr_{0.2}NiO_2$ [118] |
| $Nd_{0.8}Sr_{0.2}NiO_3$ | 260°C | 1-3 h | Pyrex tube | 2 nm STO | $Nd_{0.8}Sr_{0.2}NiO_2$ [119]123] |
|  | 280°C | 4-6 h |  | 25 nm STO |  |
| $Nd_{0.8}Sr_{0.2}NiO_3$ | 290°C | 5 h | Quartz tube | no | $Nd_{0.8}Sr_{0.2}NiO_2$ [121] |
| $Nd_{0.8}Sr_{0.2}NiO_3$ | 280°C | 4 h | PLD chamber | no | $Nd_{0.8}Sr_{0.2}NiO_2$ [122] |
| $Nd_{0.8}Sr_{0.2}NiO_3$ | 350°C | 2 h | PLD chamber | no | $Nd_{0.8}Sr_{0.2}NiO_2$ [124] [125] |
| $Nd_{0.8}Sr_{0.2}NiO_3$ | 340-360°C | 80 min | PLD chamber | no | $Nd_{0.8}Sr_{0.2}NiO_2$ [126] |
| $Nd_{0.8}Sr_{0.2}NiO_3$ | 300°C | 2 h | Quartz tube | no | $Nd_{0.8}Sr_{0.2}NiO_2$ [127] |
| $(La,Sr)NiO_3$ | 240°C | 60 min | Pyrex tube | no | $(La,Sr)NiO_2$ [128] |
|  | 260°C | 60 min |  | 2 nm STO |  |

In the table above, all the reduction agents are $CaH_2$. When using vacuum glass tubes for topotactic reduction, the standard operation process is that samples are loosely wrapped by aluminum foil, and placed them with a suitable amount of $CaH_2$ powder in glass tubes. When using PLD chambers, the thin films are wrapped in aluminum foil together with $CaH_2$ powder.

It is worth noting that most of the infinite-layer superconducting samples are under 10 nm thick. Lee *et al.* claimed that the consistency of the structure is destroyed and a secondary phase appears for thicker films, as shown in **Figure 18**a-d.[113] As the infinite-layer phase favors to appear at interfacial regions, it is reasonable to imagine that the epitaxial strain exerted by substrates may likely play important roles for chemical reduction kinetics. This could imply the importance of a capping layer.

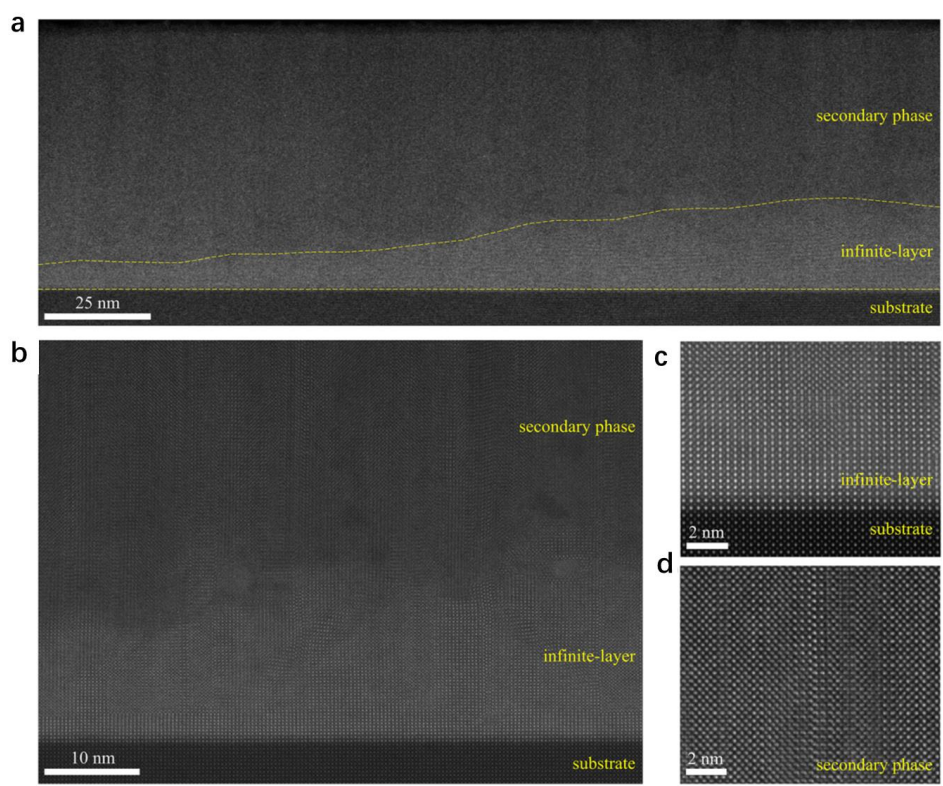

**Figure 18.** HAADF-STEM image of the obtained NSNO thin film. a) image of the mixed-phase film with substrate after the reduction. b) Magnified image of panel (a) with two phases. c) Magnified image of panel (b) in the infinite-layer region. d) Magnified image of panel (b) in the secondary phase region. Reproduced with permission.[113] Copyright 2020, American Institute of Physics.

## 5. Relevant Applications

The importance of the topotactic transition not only lies on its ability to obtain metastable phases, but also owns to numerous applications, which include superconductors, memory devices, energy storage, gas sensors, solid oxide fuel cells, batteries, catalyst and so on. In this section, we summarize some main applications closely related with the topotactic transition.

### 5.1. Superconductivity

New superconductors or higher superconducting transition temperatures of known superconductors have been always holy grails in materials science. The topotactic transition has been proved to be a useful tool for synthesizing exotic superconducting nickelates. Since the sensational discovery of the superconducting transition in the La-Ba-Cu-O system with a $T_c$ of ~30 K,[129] the non-Cu-based oxides with similar electronic structures have been long-pursing research hotspots as they may help us understand the mechanisms behind. $Ni^+$ ions in $NdNiO_2$ possess the same electronic configuration of $3d^9$ as $Cu^{2+}$ possess in $YBa_2Cu_3O_7$, which makes $NdNiO_2$ a promising candidate for novel superconductivity. However, $NdNiO_2$ could hardly be prepared using the conventional solid phase reaction method because of the relatively unstable nature of the infinite-layer phase.[130],[131]

Recently, it is found that the topotactic transition could effectively fabricate superconducting nickelate NSNO thin films, which has attracted considerable research interests. Li *et al.* firstly prepared the PV-phase $Nd_{0.8}Sr_{0.2}NiO_3$ thin films using pulsed laser deposition and then employed the topotactic chemical reduction method to convert the PV-phase to the infinite-layer phase NSNO utilizing $CaH_2$ powder as reduction agent in a Pyrex tube.[16] As shown by the lower purple line in **Figure 19**a, the temperature-dependent resistivity presents the superconducting transition with a $T_c$ of ~15 K.

Following this work, Zeng *et al.* successfully synthesized NSNO superconducting thin films using the similar topotactic transition approach.[118] Instead of utilizing a Pyrex tube, they used a pulsed laser deposition chamber to conduct the topotactic chemical reduction. In **Figure 19**b, the superconducting transitions of nickelates with different Sr doping levels from 0.15 to 2.2 can be clearly observed. Our group also demonstrated the successful preparation of a series of NSNO superconducting heterostructures utilizing this method and observed the antiferromagnetism in infinite-layer NSNO superconductors.[125] **Figure 19**c shows the superconducting transition of a NSNO/STO heterostructure with an onset transition temperature of ~13.4 K.

Stimulated by the research of NSNO, various superconducting nickelates without Nd or Sr elements have been efficaciously fabricated by the topotactic transition. For example, in 2021,

Osada *et al.* reported the synthesis of superconducting $Pr_{0.8}Sr_{0.2}NiO_2$ thin films with $T_c$ of 7-15 K as shown in **Figure 19**d.[116] One year later, Li and his coworkers produced superconducting $La_{1-x}Sr_xNiO_2$/STO heterostructures with onset transition temperatures around 9 K (**Figure 19**e).[128] More recently, Zeng *et al.* prepared $La_{1-x}Ca_xNiO_2$ thin films with a $T_c$ of ~7 K, (**Figure 19**f).[132] All the successful fabrications mentioned above confirm the effectiveness of the topotactic transition for synthesizing infinite-layer superconducting nickelates.

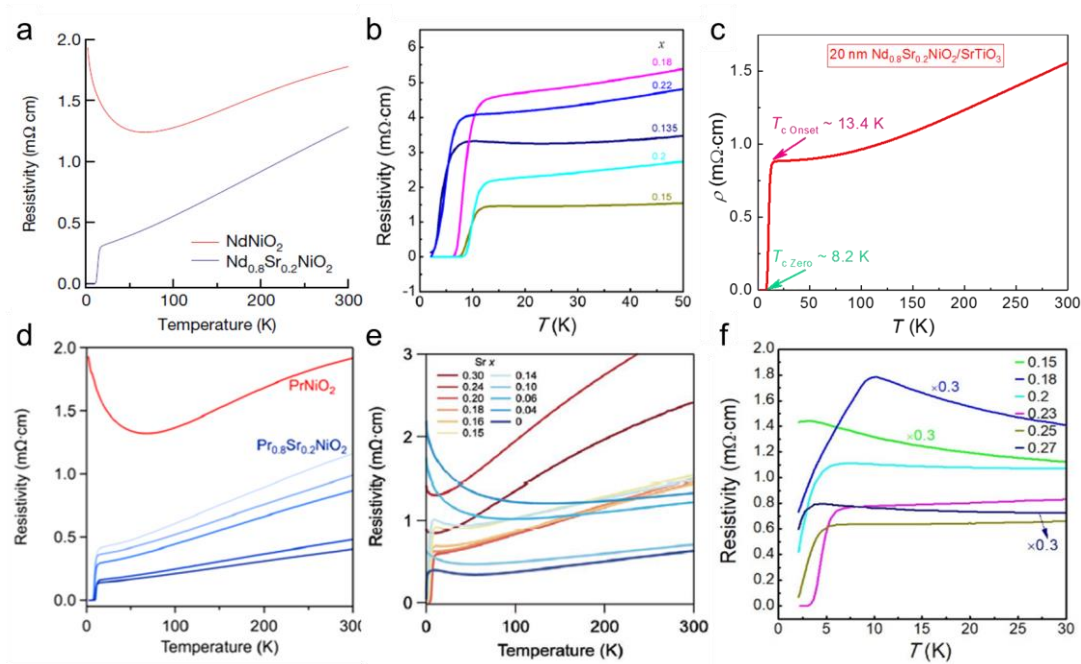

**Figure 19.** Superconductive properties of different Ni-based thin films. a) Temperature-dependent resistivity (*R-T*) with superconducting transition in NSNO system. Reproduced with permission.[16] Copyright 2019, Springer Nature. b) and e) *R-T* curve of NSNO thin film with different amount of $Sr^{2+}$ doping. Reproduced with permission.[118] Copyright 2020, America Physical Society. Reproduced with permission.[128] Copyright 2021, Wiley-VCH. c) Superconducting transition of 20nm-thick NSNO thin film grown on STO substrate. Reproduced with permission.[125] Copyright 2022, Wiley-VCH. d) *R-T* curve of $PrNiO_2$ and $Pr_{0.8}Sr_{0.2}NiO_2$ thin films. Reproduced with permission.[116] Copyright 2020, American Chemical Society. f) *R-T* curve of $La_{1-x}Ca_xNiO_2$ thin film with various $Ca^{2+}$ content. Reproduced with permission.[132] Copyright 2022, American Association for the Advancement of Science.

### 5.2. Memristive devices and artificial neural networks

A resistive switching random access memory (ReRAM) device, commonly referred to as a memristor, is deemed as an important candidate for the next-generation nonvolatile memory to replace semiconductor flash memory. The topotactic transition could be utilized to realize memristors. In recent years, new memristive devices with novel properties and high performance have emerged, based on the reversible topotactic transition between the insulating BM phase and the conductive PV phase of transition metal oxides.

In 2014, Jung's group reported the bipolar resistive switching behavior in an epitaxial (001)-oriented BM-SCO thin film grown on an STO substrate.[133] It was suggested that this phenomenon is caused by the topotactic transition of the SCO film between the BM and PV phase induced by the migration of oxygen ions excited by external bias voltages. Subsequently, in 2016, the same group achieved an SFO resistive switching device via fabricating an epitaxial (001)-oriented $SrFeO_{2.5}$/SRO heterostructure on an STO substrate.[72] This device showed remarkable bipolar resistance switching curves with a high on/off ratio, which was interpreted to be derived by the local, reversible topotactic transition of the SFO film (**Figure 20**a). Specifically, a positive bias voltage leads to the diffusion of oxygen ions from the SRO layer into the BM-SFO layer and the formation of the conducting filament consisting of the PV-SFO structure. In contrast, a negative bias voltage results in the rapture of the conducting filament as shown in **Figure 20**b-d.

Although the resistive switching devices based on the BM metal oxide films were built, the direct observation of the topotactic transition during the resistive switching process was missing. Later in 2019, Tian *et al.* straightforwardly observed the formation, extension and rupture of the PV-SFO nanofilament through the BM-SFO layer in an epitaxial SFO/SRO heterostructure by scanning transmission electron microscopy (STEM), **Figure 20**e,f.[75] This work clearly revealed that the resistive switching is induced by the topotactic transition mechanism with the moving and ordering of the oxygen ions. Almost at the same time, Jung's group observed the topotactic transition process that gives rise to the formation and rupture of the nanofilament in an SFO-based ReRAM device via x-ray absorption spectroscopy (**Figure 20**g-l), which provided more evidence for the topotactic transition based resistive switching behavior in this class of memristors.[73] The topotactic transition dominated resistive switching were subsequently demonstrated in other works (**Figure 20**m).[74] More recently, in 2022, Lo *et al.* grew an epitaxial CaFeO$_{2.5}$ thin film on a Nb-STO substrate and observed the resistive switching in a top electrode/CaFeO$_x$/Nb-STO heterostructure (**Figure 20**n).[134] Similarly, it was found that the PV-CaFeO$_3$ nanofilament was formed and ruptured in the BM-CaFeO$_{2.5}$ layer through the topotactic transition accompanied by the resistive switching behavior using STEM. Moreover, they studied the valence state of Fe in CaFeO$_x$ and found that Fe$^{4+}$ ions were present in the localized PV-CaFeO$_3$ structure via electron energy loss spectroscopy analyses, which proves the formation of the PV-CaFeO$_3$ phase.

This new class of resistive switching devices, in which the resistive switching is attributed to the formation and rupture of the conductive PV nanofilament caused by the reversible topotactic transition between the BM and PV phase, is different from traditional transition metal oxide memristors based on the oxygen vacancy rich filament. These findings also point out a new direction for the performance optimization of ReRAM. For instance, the size of the PV nanofilament is ultra-small, which could be beneficial to develop highly packed nanoscale ReRAM. In addition, these devices also exhibit large ON/OFF ratios, fast switching speeds, high endurance cycles and long retention time, which offer great advantages for nonvolatile memory device applications.

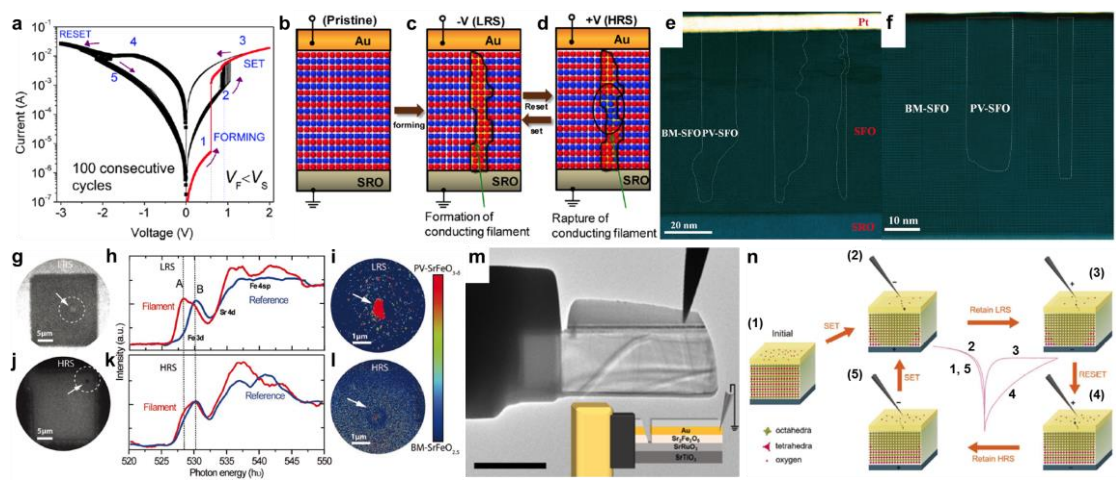

**Figure 20.** a) Repeatable resistive switching curves the SFO film. b-d) Illustrative diagram of the topotactic transition process of the (001) SFO device in the pristine state (b), low-resistance state (LRS) (c) and high-resistance state (HRS) (d). Reproduced with permission.[72] Copyright 2016, American Chemical Society. e,f) STEM image of the PV-SFO nanofilaments through the BM-SFO film in the set state (e) and post-reset state (f). Reproduced with permission.[75] Copyright 2019, Wiley-VCH. g-i) Photoemission electron microscope image (g,j), O K-edge spectra (h,k) and false-color contribution mapping of the PV-SFO and BM-SFO (i,l) of the SFO (111) device in the LRS (g-i) and HRS (j-l). Reproduced with permission.[73] Copyright 2019, Wiley-VCH. m) TEM image of the SFO device structure. Reproduced with permission.[74] Copyright 2020, AIP Publishing. n) diagrammatic drawing of the topotactic transition process accompanied by the resistive switching in the $CaFeO_x$ device. Reproduced with permission.[134] Copyright 2022, Wiley-VCH.

Besides the nonvolatile memory applications, memristive devices are one of the most promising candidates for neuromorphic circuits. The topotactic transition-based ReRAM could be utilized to implement memristive synaptic devices and neuromorphic computing. In 2019, Ge *et al.* reported a three-terminal synaptic transistor based on an SFO memristor, as shown in **Figure 21**a,b, in which the topotactic transition between the BM-SFO and the PV-SFO phase could be controlled by the electrical voltage applied on the gate of the transistor and then led to the gate-controllable multilevel conduction states.[77] This artificial synaptic device could mimic the functions of human brain synapses. Remarkably, the artificial neural network composed of this kind of synaptic transistors shows accurate (95.2%) classification of handwritten data, **Figure 21**c,d. Then in 2020, Jung's group achieved a synaptic-like memory device on the base of an SFO (111) topotactic transition memristor, which displays synaptic-like operations and exhibits a recognition accuracy of ~90% (**Figure 21**e,f).[76] Furthermore, in 2021, Mou *et al.* demonstrated an SCO-based topotactic transition memristor.[46] They used these devices to mimic an artificial neural network, which could not only pruned the neural network by ~84.2% but also enhanced the classification accuracy to 99% **Figure 21**g-i). Recently, in 2022, Chen *et al.* successfully established topotactic transition-based artificial synapses and neurons consisting of BM-SFO/SRO and PV-SFO/BM-SFO/SRO memristors, respectively.[135] They built fully memristive neural networks using these artificial synapse and neuron devices, which exhibit good performance on unsupervised image recognition and an accuracy up to 100% for recognizing images (**Figure 21**j-n). Overall, the topotactic transition paves a brand-new way to designing artificial neural networks and implements high-performance neuromorphic computing.

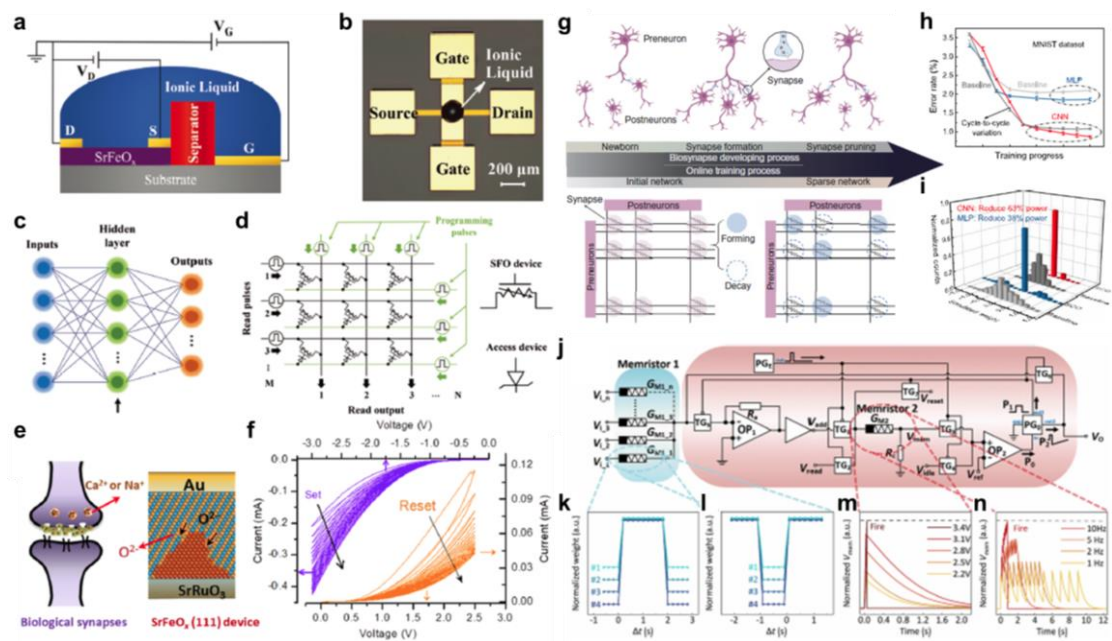

**Figure 21.** a) Schematic of the three-terminal synaptic transistor. b) Optical photo of the synaptic transistor device. c,d) Illustrative diagram of the artificial neural network (c) and hardware implementation (d) consisting of the synaptic transistor. Reproduced with permission.[77] Copyright 2019, Wiley-VCH. e) Sketch diagram of the biologic synapse and the SF$_x$ (111) device. f) Resistive switching curves of the SFO (111) device. Reproduced with permission.[76] Copyright 2020, American Chemical Society. g-i) Schematic of the online training (g), the recognition error rate (h) and the simulated synaptic weight distribution (i) of the artificial neural network composed of the SCO-based TPT-RAM synapses. Reproduced with permission.[46] Copyright 2021, American Association for the Advancement of Science. j) Illustrative diagram of the fully memristive neural network comprised of the SFO-based artificial synapse and neuron devices. k,l) Simulated spiking-timing dependent plasticity curves of the artificial synapse devices with the HRS (k) and LRS (l). m,n) Simulated leaky integration and fire behaviors of the artificial neuron devices with the spatial integration (m) and temporal integration (n). Reproduced with permission.[135] Copyright 2022, Elsevier.

## 5.3. Fuel cells and batteries

Nowadays, environmental protection has become a key topic of human society and clean energy is growing to take place of traditional fossil fuels gradually, especially for the automobile industry. The booming new energy transports are mainly driven by electricity at the current stage, which is in urgent need of innovative approaches or new materials for electrolytes and electrodes to realize larger capacity densities and higher conversion efficiencies. And it is no doubt that in our daily life, industrial manufacture and many other fields could also benefit from this technology development, such as stationary generators and portable power systems.

As a type of fuel cells converting chemical energy to electric one, solid oxide fuel cells (SOFCs) possess the capability of fuel flexibility and overcoming combustion efficiency limitations.[136] However, SOFCs rely heavily on the ion diffusion and oxygen reduction reaction and the primary bottleneck issue is the high operating temperature.[137],[138] As a consequence, redox-active topotactic transition metal oxides with the PV structure, such as SCO[81],[85],[88],[90],[93] and SFO[70],[139],[140], are proposed to be potential candidates for SOFCs, owing to the nature of relatively high ionic and electronic conductivities at low temperatures. For example, $La_{0.5}Sr_{0.5}Co_{0.5}Ti_{0.5}O_{2.64}$ (LSCTO) has been synthesized by a topotactic reduction process from the oxidized PV phase.[141] When LSCTO works as a cathode, it shows good performance and achieves output power densities close to 500 mW/cm$^2$ at 850°C. At intermediate temperatures, the products of the topotactic transition can still yield 220 mW/cm$^2$ as a cathode or 170 mW/cm$^2$ as an anode, respectively. Similarly, double PV-$Sr_2ScTi_{0.5}Mo_{0.5}O_6$ evolved from the cubic $ScTiO_3$ due to the topotactic oxidation and exhibited improved electrode performance of ~220 mW/cm$^2$ at 800 °C in SOFCs.[142] In addition, Zhuang *et al*. demonstrated that acidic $MoO_3$ can capture SrO from $(La_{0.6}Sr_{0.4})_{0.95}Co_{0.2}Fe_{0.8}O_{3-\delta}$ (LSCF) to transform topotactically into $SrMoO_4$.[143] During this process, more Sr/O vacancies form at the LSCF surfaces (**Figure 22**a) and the remaining octahedra acquire more structural flexibility. Eventually, the redox activity of oxygen in the lattice is increased. As shown in **Figure 22**b, the oxygen redox activity of LSCF greatly enhances the cathode performance for SOCF at intermediate temperatures. More importantly, the elimination of strontium segregation and the formation of electrochemically inerted SrO islands extend the service life of the cathode.

Besides, proton exchange membrane fuel cells are the major fuel cells around the world. The topotactic transition, serving as a route towards novel materials, has also attracted broad interest in this area. For instance, Pt/RuO$_2$ nano structures have been topotactically transformed from an icosahedral Pt/Ni/Ru nanocrystal by Oh *et al.* (**Figure 22**c).[144] They observed rattle-like Pt/RuO$_2$ showing outstanding activity and high durability for proton exchange membrane electrolyzers. Double-layer Pt nanosheets can also be synthesized directly through the topotactic transition of single-layer platinum oxide nanosheets (**Figure 22**d).[145] The topotactically reduced Pt nanosheets with a thickness of 0.5 nm have larger electrochemically active surface areas and the resulting greater performance. It could reduce the usage of Pt and lower the production cost of proton exchange membrane fuel cells. In a similar way, ultrathin Bi nanosheets were realized by the *in-situ* topotactic transition from BiOI[146] and BiOCl[147] nanosheets as well. The products with the advantages of single crystallinity and enlarged surface areas exhibit excellent durability and performance, which are favorable for energy conversion.

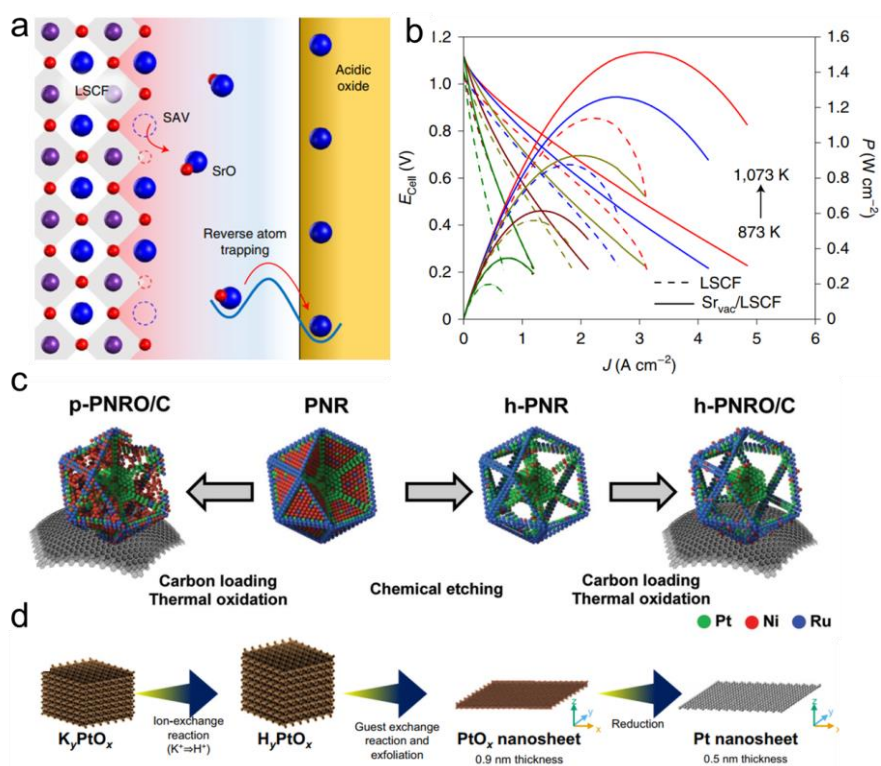

**Figure 22.** Topotactic transition for application of fuel cells. a) Schematic of SrO captured by acidic MoO$_3$ to transform topotactically to SrMoO$_4$. b) The performance of SOFCs with LSFC and strontium-deficient LSCF. Reproduced with permission.[143] Copyright 2022, Springer Nature. c) Schematic illustration of the topotactic transition for fabricating Pt/RuO$_2$

nanostructures. Reproduced with permission.[144] Copyright d) Diagram of the synthetic approach of double-layer Pt nanosheets. Reproduced with permission.[145] Copyright 2020, Elsevier.

It is worth noting that apart from fuel cells, the topotactic transition can likewise play an important role in ion batteries. In 2013, Nakajima *et al*. obtained $LiCoO_2$ mesocrystals by the topotactic transition and the products exhibited enhanced charge-discharge cycle stability and rate performance in lithium-ion batteries.[148] Similar topotactic transitions of Li-Mn-O mesocrystals from $MnCO_3$ were achieved almost simultaneously.[149] Most recently, Jia *et al*.[150] proposed that the rock salt/spinel phase on the cathode surfaces of lithium-ion batteries could undergo the topotactic transition into $Ni_{0.5}Co_{0.2}Mn_{0.3}(OH)_2$, and finally back to $LiNi_{0.5}Co_{0.2}Mn_{0.3}O_2$ (**Figure 23**a). Compared with non-topotactic processes, the topotactic transition helps to form better $Li^+$ transport channels and improves the efficiency of circulation for spent cathode materials. In addition to lithium oxides mentioned above, rutile $TiO_2$ hollow nanocrystals[151] and $Co_3O_4$ nanotubes[152] fabricated via the topotactic transition could act as lithium-ion battery electrodes. In terms of sodium-ion batteries, $Na_xCoO_2$ epitaxial thin films could be synthesized by combining molecular-beam epitaxy and the topotactic transition to serve as a potential cathode material (**Figure 23**b).[153] Additionally, $Na_xVO_2$[154] and other binary systems $Na_x(Mn,Ni)O_2$,[155] $Na_x(Fe,Co)O_2$,[156] or $Na_x(Fe,Ni)-O_2$[157] have also been reported through the topotactic process for the application of sodium batteries.

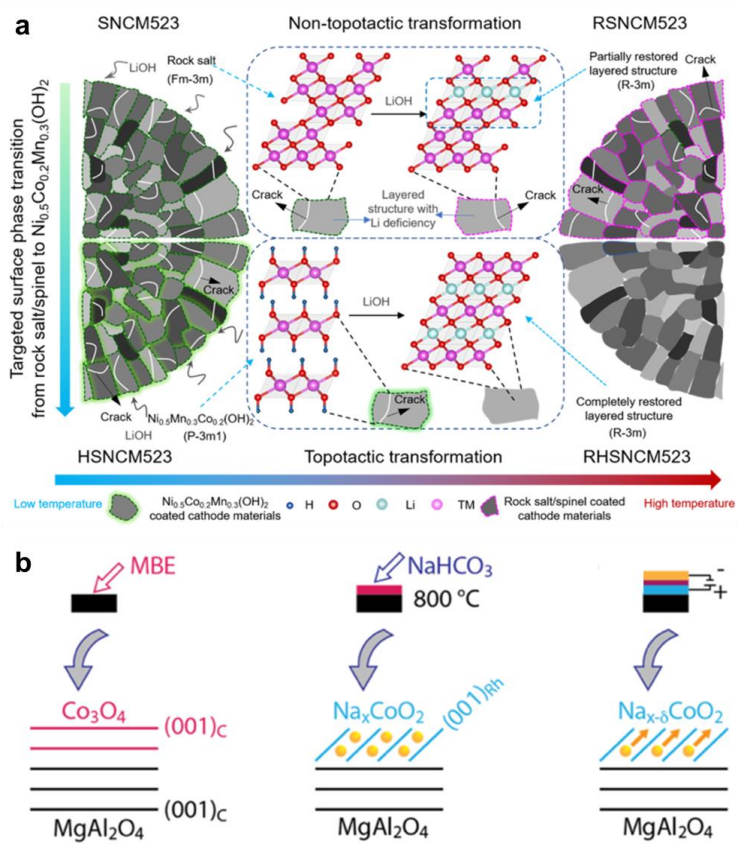

**Figure 23.** The application of the topotactic transition in ion batteries. a) Mechanisms for the regeneration of LiNi$_{0.5}$Co$_{0.2}$Mn$_{0.3}$O$_2$ via topotactic transition. Reproduced with permission.[150] Copyright 2023, American Chemical Society. b) Schematic of the synthesis process of Na$_x$CoO$_2$ through the topotactic transition. Reproduced with permission.[153] Copyright 2023, American Chemical Society.

## 6. Summary and Outlook

In this review, we have introduced an important route to the birth of new oxide materials: topotactic transition, which largely differs from usual chemical doping but is characterized by collective ordered gain(loss) and rearrangement of atoms leading a relating structural phase transition. To have a deeper understanding of the attractive topotactic transition, the structural-facilitated effect, the mechanism of oxygen migration kinetics, a variety of technical viable methods for different types of metal oxides as well as several important potential applications are comprehensively summarized. This work is to provide straightforward guides for understanding the structural natures, kinetic mechanisms and new routines to new oxide materials with unpredictable chemical, physical, and even biological functions.

Up to now, the topotactic transition has been only regarded as a novel approach to new materials in various studies. There are investigations on the structure and kinetic mechanisms of topotactic transitions yet it is still not enough. It would be rather insightful if advanced time- and space-resolve structural characterization even some isotopic labeling techniques could be utilized to more studies to unveil the key factors that are pivotal for determining the kinetic processes of topotactic transitions, for example, strain in thin films, defects in thin films, chemical affinities of key metal atoms to oxygen, diffusion coefficients of atoms and so on. Additionally, besides the PV phase, the BM phase and the infinite-layer phase, more feasible pristine structures favorable for the topotactic transition are to be explored to enlarge the material scope of the topotactic transition.

With the fast development of materials science and especially the strong and broad application demand of oxides in energy storage materials, energy conversion materials, catalysts, information technology and other areas, we hope to see more and more breakthroughs in terms of the emergence of new materials achieved via the topotactic transition as well as the studies on the dominant driving force of this novel structural transition itself.

**Acknowledgements**
Z.L. acknowledges the financial support of the National Key R&D Program of China (Grant No. 2022YFB3506000). Z.L. acknowledges the financial support of the National Key R&D Program of China (Grant No. 2022YFA1602701). Z.L. acknowledges the funding supported by the National Natural Science Foundation of China (Nos. 52271235 and 52121001).